# Critical Cooperation Range to Improve Spatial Network Robustness

V. H. P. Louzada,[1, *] N. A. M. Araújo,[2, 3] T. Verma,[1] F. Daolio,[4] H. J. Herrmann,[1, 5] and M. Tomassini[4]

[1]*Computational Physics, IfB, ETH Zurich, Wolfgang-Pauli-Strasse 27, 8093 Zurich, Switzerland*
[2]*Departamento de Física, Faculdade de Ciências,*
*Universidade de Lisboa, P-1749-016 Lisboa, Portugal*
[3]*Centro de Física Teórica e Computacional, Universidade de Lisboa,*
*Avenida Professor Gama Pinto 2, P-1649-003 Lisboa, Portugal.*
[4]*Faculty of Business and Economics, University of Lausanne, Lausanne, Switzerland*
[5]*Departamento de Física, Universidade Federal do Ceará, 60451-970 Fortaleza, Ceará, Brazil*

(Dated: September 4, 2018)

A robust worldwide air-transportation network (WAN) is one that minimizes the number of stranded passengers under a sequence of airport closures. Building on top of this realistic example, here we address how spatial network robustness can profit from cooperation between local actors. We swap a series of links within a certain distance, a cooperation range, while following typical constraints of spatially embedded networks. We find that the network robustness is only improved above a critical cooperation range. Such improvement can be described in the framework of a continuum transition, where the critical exponents depend on the spatial correlation of connected nodes. For the WAN we show that, except for Australia, all continental networks fall into the same universality class. Practical implications of this result are also discussed.

## INTRODUCTION

The construction of a new terminal at the Schenzen airport, Southeast China, has been used to question the current strategies of infrastructure growth in developing countries [1, 2]. Schenzen is a large city, but its airport is directly connected by an eight kilometer ferry to the Hong Kong airport, which can handle twice as many passengers. Is it reasonable to invest more than one billion dollars increasing the capacity of such a large infrastructure with another one nearby? Opponents to this investment classify it as a white elephant and as one example of the misbegotten infrastructure growth in developing countries. The ones in favor, argue that not only the costs of investing in Schenzen are lower than in Hong Kong but also global connectivity can profit from cooperation between stakeholders of air-transportation systems in the region. Here we address, from a network science perspective, how such proximity and cooperation between local actors is key to the robustness of spatial networks.

The simplified representation of complex systems as a network of nodes and links has provided important insights into the design of a variety of systems, such as power grids [3–5], maritime commerce [6], and communication networks such as the Internet [7–9]. In many cases, nodes are spatially embedded according to the spatial coordinates of the elements [10, 11]. This simplification allows us to focus on the topological aspects of the system and to easily extend our results to many applications. We characterize the robustness of a network as its capacity to maintain global connectivity under a sequence of node removals and describe a strategy based on local cooperation to improve robustness under possibly realistic constraints. We discover a continuum transition when changing the distance for which nodes are allowed to swap links, a cooperation range. We calculate the critical exponents of this transition and show that the key factor controlling the value of the critical exponents is the exponent of the algebraic decay of the connection probability with the node distance.

As an ubiquitous infrastructure system, we explain our method in the context of the worldwide air-transportation network (WAN), though our results impact the whole class of spatially embedded networks. It is paramount that the WAN works in an extremely reliable and efficient fashion, as any temporary airspace closure, such as the one caused by the eruption of the volcano Eyjafjallajökull in 2010, may cause huge losses worldwide [12, 13].

Complex Networks have been used to study airflight networks. Simple abstractions of flights and airports have been used to characterize its robustness [14, 15], study its structural properties [16–25] and evolution [26, 27]. Here instead we go one step further and propose modifications to the robustness of the WAN and identify new properties associated with it. We summarize data provided by OpenFlights in the year 2011 as a single static network with 3237 airports (nodes, modeled as points distributed across the surface of a sphere with distance calculated according to the Haversine formula) and around eighteen thousand links [28]. Links are undirectional, assuming that each flight should return to its origin, and weighted according to the number of possible flights between two airports. Airports are weighted by the total number of passengers transported that year. Information regarding the number of passengers transported between airports is not available.

As a self-organized system, in which preferential attachment is expected to play a pivotal role, the WAN

is quite fragile to targeted attacks, i.e., intentional removal of the most connected nodes causing the collapse of the giant connected component [25]. The aim of our optimization strategy is to create a robust yet economically feasible WAN. Since robustness can be defined in different ways, we consider that a robust WAN should be capable of transporting passengers even in face of a targeted attack, in contrast with previous works where only the size of the largest connected component is considered [29–34]. We simulate a sequence of airport closures (node removal) and quantify robustness $r$ as:

$$r = \frac{1}{\Pi(0)} \sum_{n=1}^{N} \Pi\left(\frac{n}{N}\right) , \qquad (1)$$

where $N$ is the total number of nodes, $n$ is the number of nodes removed from the network, and $\Pi(q)$ is the number of passengers in the largest component after a fraction $q = n/N$ of nodes were removed, i.e., the sum of nodes' weight on the largest component. Closures are executed from the most to the least connected node.

The location of airports are mostly determined by economical forces, such as to cater to local demands. In many cases, airports are located within a short distance from each other, sometimes only a few kilometers away as, e.g., airports in the Schenzen-Hong Kong area, or a few hundred kilometers but still easily reachable, such as the airports in the northeast of the United States. We assume that a flight rerouted to an airport within a cooperation range $v$ of the original destination has a similar attractiveness. If need be, a passenger landing at a different airport could easily take another means of transportation, such as the local train network or a shuttle bus, to go to the desired destination. For transportation networks, the cooperation range is defined as the geographical distance between nodes, but other spatial networks might require other metrics, such as travel time or cost.

We increase network robustness through link swaps, with the probability to swap a route inversely proportional to the weight of a link $e_{ij}$ between airports $i$ and $j$, so that important connections are affected with less priority. This tends to keep the transportation capacity of the system stable, and the introduction of complicated interventions on airports, such as building new runaways, might be avoided. Moreover, a connection is only rerouted to an airport within the cooperation range of the original destination. As an example, flights could be distributed between Hong Kong and Schenzen or among the airports surrounding London.

Given a cooperation range $v$ and a metric $d(i, j)$, which calculates the distance between nodes $i$ and $j$, the following swap strategy is performed:

1. Select a node $i$ randomly having at least one neighbor;

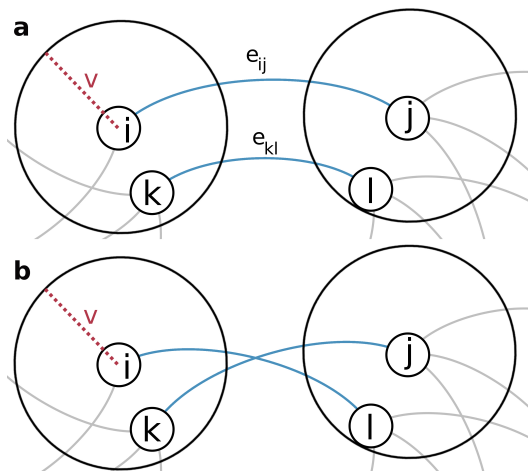

FIG. 1. **Geo swap: a cooperation range rewiring.** Diagrammatic representation of a rewiring procedure based on a cooperation range. For a randomly selected node $i$, a node $k$ at distance $d(i,k) \leq v$ is selected (Panel a). If nodes $j$, neighbor of $i$, and $l$, neighbor of $k$, also have $d(j,l) \leq v$ then links $e_{ij}$ and $e_{kl}$, in blue, are swapped (Panel b).

2. Select a neighbor $j$ of node $i$ with probability inversely proportional to the weight of $e_{ij}$;

3. Select a pair of connected nodes $k$ and $l$ so that $d(i,k) \leq v$ and $d(j,l) \leq v$ with probability inversely proportional to the weight of $e_{kl}$;

4. Remove links $e_{ij}$ and $e_{kl}$ and create links $e_{il}$ and $e_{jk}$.

This strategy is illustrated in Fig. 1. Swaps change the network robustness $r$ but we only perform swaps that increase $r$. From this point on, we call such a swap, a *geo swap*. A fixed number of geo swaps of the order of the number of links is executed. To compare networks of different sizes and populations, we normalize the robustness as $R = (r - r_{\min})/(r_{\max} - r_{\min})$, in which $r_{\min}$ is the value of $r$ for $v = 50$ km, the minimum value for which a geo swap will be considered in the WAN, and the maximum robustness $r_{\max}$ obtained for $v = 18 \times 10^3$ km, which is approximately half of the planet perimeter. This strategy builds on top of previous work which focused on different acceptance mechanisms [30, 33] or topological characteristics [34], but differs significantly by focusing on geographic limitations and low-weight links.

## RESULTS

The cooperation range $v$ limits the area of possible swaps to guarantee that a geographically acceptable change is performed. A too small value of $v$ does not provide sufficient room for robustness improvement. While a too large value of $v$ leads to reroutes of connections to impractically far away airports. By tuning the values of $v$, we observe a critical value of the cooperation range $v^* = 910 \pm 90$ km at which a significant improvement in the WAN is first registered. This range in fact yields the highest variance of robustness increase among all possibilities, as shown in Fig. 2.

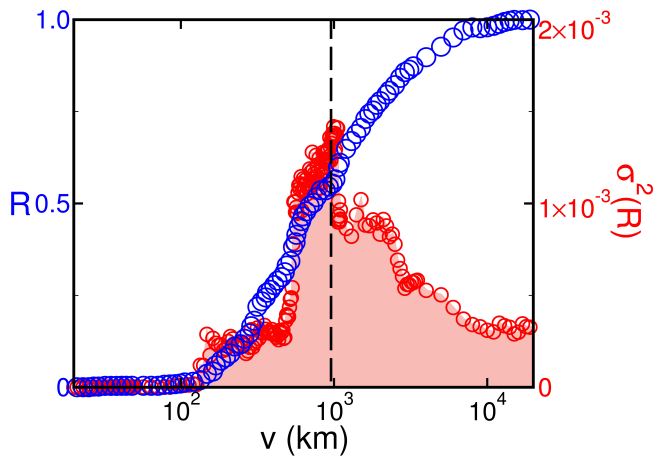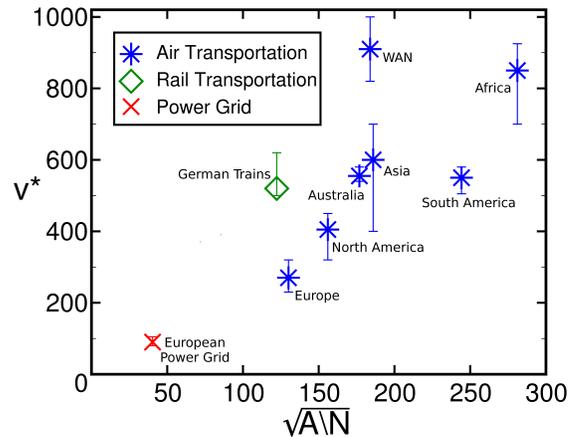

FIG. 2. **Robustness increase of the WAN.** Dependence of the optimized robustness on the cooperation range (blue) and variance over samples (red) with a maximum at $910 \pm 90$ km, in which the standard deviation reachs the maximum.

FIG. 3. **Critical cooperation range as a function of the nodes' coverage.** The critical cooperation $v^*$ is positively correlated with $\sqrt{A/N}$, where $A$ is the total area in which $N$ nodes are embedded. The three different symbols represent the types of infrastructure networks in which the geo swap is applied.

By considering continental airflight networks - constructed from the division of the WAN into continents (details in **S1** Table) - together with other spatially embedded networks (the European Power Grid and the European Rail network), we observe that $v^*$ is positively correlated with $\sqrt{A/N}$ (Spearman's correlation coefficient $\rho = 0.78$), where $A$ is the total area in which the $N$ nodes are embedded (Fig. 3). Being the combination of local and intercontinental flights, the WAN lies slightly off the trend, but in general we can conclude that the typical radius served by an airport is correlated to the minimum distance at which swaps become effective. Artificially generated random networks also confirm this relationship (**S1** Figure). Other topological characteristics of the optimized networks are detailed in **S4** Figure.

Close to the critical cooperation range, the evolution of $R$ scales with $v - v^*$ for the continental networks. Applying the finite-size scaling,

$$R = N^{-\frac{\beta}{\nu}} \mathcal{F}\left[(v - v^*) N^{\frac{1}{\nu}}\right], \quad (2)$$

where $\nu$ and $\beta$ are critical exponents and $\mathcal{F}[x]$ is a scaling function, we collapse the data for different system sizes. For $v = v^*$, as shown in the inset of Fig. 4, $R$ scales with $N^{-\frac{\beta}{\nu}}$, with $\beta/\nu = 0.08 \pm 0.01$, as expected for a continuous transition. This allows us to calculate the exponents in the main panel of Fig. 4 as $\beta = 0.20 \pm 0.02$ and $1/\nu = 0.40 \pm 0.03$. The data suggest that the construction of airports and the creation of connections follow a similar mechanism in all continents, though the limited system size of each continent and obvious geographic differences prevent strong conclusions. However, data for Australia significantly differs from the others. Because a great number of islands in Oceania have many small airports, sometimes being the only feasible connection between remote areas, we assume that airports and flights in this continent were established following a different mechanism. Most probably, the predominance of several small islands might pose a physical limit to the cooperation range which is related to the average distance between islands.

Spatial networks are mainly defined by three properties: nodes' position, degree distribution, and connection pattern. A simplified model - where nodes are assigned random positions, the degree distribution is a Poisson distribution, and connections are randomly assigned without any spatial/degree bias - displays different critical exponents (**S3** Figure). However, if the probability $P(i,j)$ that nodes $i$ and $j$ are connected decays algebraically with the distance between $i$ and $j$,

$$P(i,j) \propto \frac{1}{d(i,j)^\alpha}, \quad (3)$$

where $\alpha \in \mathbb{R}$ is the decay exponent, we obtain exponents that are numerically consistent with the ones in Fig. 4. Based on a simplification of the gravity model, used to describe connections between geographically distributed nodes [6, 35–37], we call Eq. (3) a *distance-decay* model as it does not take into account the degree/weight of the nodes to calculate $P(i,j)$.

To test Eq. (3), we plot a new data-collapse in Fig. 5 based on networks generated as follows. Node positions are uniformly distributed across an Earth spherical cap of area $A$ and in order to keep $\sqrt{A/N} \approx 195$, the same value of the WAN, network size is calculated accordingly. Node degree follows a Poisson distribution of average degree 12. Node weights are chosen according to the equation $W(i) = 102.6 k_i^{1.1}$, where $k_i$ is the degree of node $i$, which is a fit of the relationship between node weights and degree of the WAN. Link weights are randomly distributed from $[1, 14]$, in which 14 is the maximum link weight on the WAN. We observe that changes in the value of $\alpha$ affect consistently the slope in the data-collapse (Fig. 5**b**). We find a value of $\beta$ similar to the one of the continents, without Australia, for $\alpha \in [1.8, 2.0]$. For $\alpha = 2.0$, the finite-size scaling in Fig. 5c allows us to estimate: $\beta = 0.23 \pm 0.02$ and $\beta/\nu = 0.08 \pm 0.01$ (Fig. 5**a** and Fig. 5**c**). Further analysis also show that when $\alpha \approx 2$ the ratio between the average length of routes and the average distance between two airports is similar to that found for the continents (**S2** Figure). Interestingly, the empirical probability distribution of link lengths in the WAN is a power law of exponent $\alpha = 2.2 \pm 0.2$ (**S5** Figure). This suggests that correlations as the ones developed in the distance-decay model are consistent with the ones found for the WAN and characteristic for the universality class.

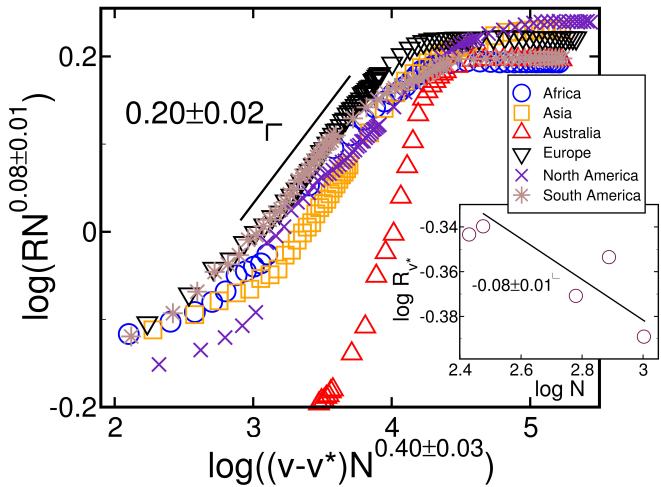

FIG. 4. **Continuous transition of robustness.** Data collapse of the robustness evolution for $v-v^* > 0$ after successive applications of the geo swap. Curves in the main panel represent each continent. A total of $10^4$ tentative geo swaps are executed for several cooperation range values. The value of $v^*$ is selected as the highest variance point over 100 samples. Data is scaled using $1/\nu = 0.40$ and $\beta/\nu = 0.08$. The inset shows size dependence of $R$ at $v = v^*$, scaling as $R \sim N^{-\frac{\beta}{\nu}}$, with $\beta/\nu = -0.08 \pm 0.01$, where $N$ is the total number of nodes. Symbols are larger than the standard deviation.

## DISCUSSION

In order to provide applicable insights, any network modification strategy should take into account realistic constraints naturally imposed by the problem. The geo swap contains a simple set of rules specifically designed to improve the robustness of spatial networks. It is important to note however that a probabilistic approach is more a guidance than a closed optimization recipe. We expect that future procedures built on top of our strategy should be carefully tailored to the underlying system.

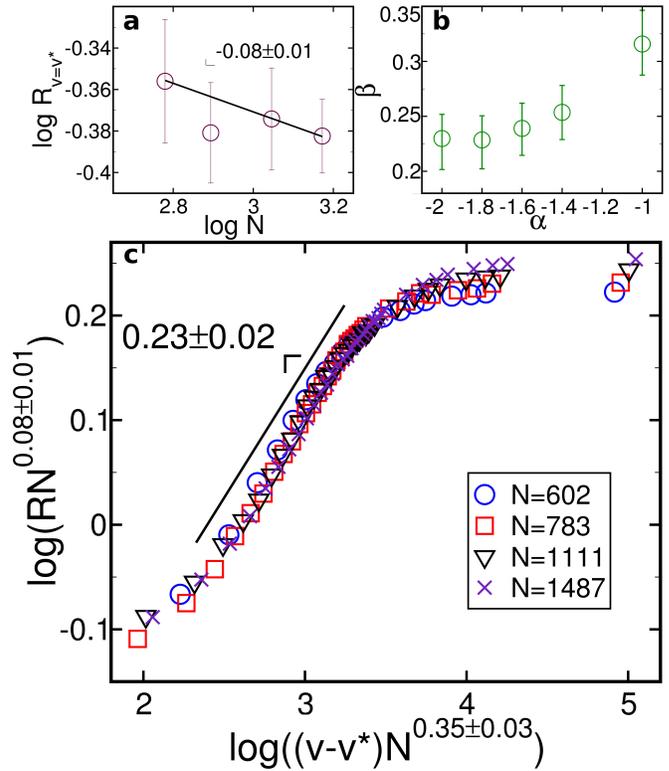

FIG. 5. **Distance-decay model reproduces the same critical exponents of the WAN.** Data-collapse of the robustness evolution for $v-v^* > 0$ after successive applications of the geo swap on random networks generated through the distance-decay model with $\sqrt{A/N} \approx 195$. **a**, value of $R$ at $v = v^*$ with $\alpha = 2.0$ scaling with the number of nodes $(N)$ as $R \sim N^{-\frac{\beta}{\nu}}$, with $\beta/\nu = -0.08 \pm 0.01$. **b**, impact on $\beta$ for the distance-decay model with different values of $\alpha$. **c**, curves for different system sizes with $\alpha = 2.0$, scaled using $1/\nu = 0.35$ and $\beta/\nu = 0.08$.

In an infrastructure perspective, the geo swap, which makes flights land in different airports, has two crucial implications. Firstly, a second transportation system should be used to connect nearby airports, in line with recent works dealing with the coupling of infrastructure networks [38]. By taking into account the critical cooperation range these couplings could be designed or improved for distances close to $v^*$. Secondly, a local level of cooperation is necessary between airports. Flights in Schenzen and Hong Kong could be rerouted to attend different yet nearby airports worldwide, further increasing the reliability of the local service and the overall WAN robustness.

One could also organically extended our analysis to in-corporate airlines into the scenario. The already existing alliances within the airline industry do cooperate on a regular basis. This direction could be pursued by future works if data about alliances among airlines is available.

Our rewiring strategy is also able to show that the continents, with one exception, follow the same universality class regarding robustness improvements, as the probability that two airports are connected decays quadratically with their distance. Being a continent with its own geographical idiosyncrasies, Australia does not fit our analysis, for which further studies are necessary. In summary, our results show that, for any spatial network, the universality class of the robustness improvement strongly depends on the spatial correlation of connected nodes.

## METHODS

Figure 2 results from the application of $10^4$ tentative geo swaps for each value of $v$ in the WAN. Each blue circle is the average over 250 samples while red circles stand for the variance over samples.

In Fig. 3, data for the European Power Grid (red) is retrieved from Ref. [39] and the Rail transportation network was manually assembled using public data. The power grid network has 1254 nodes and 1812 links, and the rail network has 39 nodes and 70 links. For continents, the power grid, and the rail network, a total of $10^4$ tentative geo swaps are executed for several cooperation range values. The value of $v^*$ is selected as the highest variance point over 100 samples, with error bars representing the values where variance is equal to $0.75\sigma^2(v^*)$. The same data for the continents is used to construct Fig. 4, in which symbols are larger than the standard deviation.

For all panels in Fig. 5, a maximum of $10^4$ tentative geo swaps are executed for several cooperation ranges. Each point represents the average over 100 samples, with symbols being larger than standard deviation in Panel **c**. The critical cooperation range is defined as $v^* = 240 \pm 10$ km.

**Acknowledgments.** Authors would like to thank the CNPq, Conselho Nacional de Desenvolvimento Científico e Tecnológico - Brasil, the European Research Council advanced grant FP7-319968-flowCSS, and the ETH Zurich Risk Center for financial support. NA acknowledges financial support from the Portuguese Foundation for Science and Technology (FCT) under Contract no. IF/00255/2013. We would also like to thank I. Nikolakopoulos and K. J. Schrenk for providing the caffeine for the discussion in which the idea of this paper was conceived.

# Supplementary Material
## Critical Cooperation Range to Improve Spatial Network Robustness

V. H. P. Louzada,[1, *] N. A. M. Araújo,[2, 3] T. Verma,[1] F. Daolio,[4] H. J. Herrmann,[1, 5] and M. Tomassini[4]

[1]*Computational Physics, IfB, ETH Zurich, Wolfgang-Pauli-Strasse 27, 8093 Zurich, Switzerland*
[2]*Departamento de Física, Faculdade de Ciências,*
*Universidade de Lisboa, P-1749-016 Lisboa, Portugal*
[3]*Centro de Física Teórica e Computacional, Universidade de Lisboa,*
*Avenida Professor Gama Pinto 2, P-1649-003 Lisboa, Portugal*
[4]*Faculty of Business and Economics, University of Lausanne, Lausanne, Switzerland*
[5]*Departamento de Física, Universidade Federal do Ceará, 60451-970 Fortaleza, Ceará, Brazil*
(Dated: March 24, 2015)

## TABLE S1 - THE WORLD AIR-TRANSPORTATION NETWORK

The network that represents the WAN was carefully assembled in Ref. [1], using data from OpenFlights (http://sourceforge.net/p/openflights/code/682/tree/openflights/data/, files airports.dat and routes.dat) and The World Bank dataset sourced through Civil Aviation Statistics of the World and ICAO staff estimates (http://data.worldbank.org/indicator/IS.AIR.PSGR), all for the year 2011. After removing redundancies and extrapolating some passengers information (Ref. [1]), we construct a network of 3237 airports and 18.125 flights.

Information about continents, contained in the original data set, was used to further break it down. Continental networks contain flights in which both end-points are in the same continent. Australia includes all islands in the Pacific ocean. For simplification, Russia is entirely part of Europe, and Turkey is entirely in Asia. A summary of the main characteristics of the continents and the WAN is present in the table below.

TABLE S1. Data of WAN and its continental components.

| Name | Nodes | Links | Passengers (daily) | Flights (daily) | Average Degree | Average Passengers (daily) | Average Flights (daily) | Average Distance btw Airports (km) | Average Flight Distance (km) |
|---|---|---|---|---|---|---|---|---|---|
| Africa | 269 | 642 | 220,481 | 3,023 | $4.77 \pm 0.43$ | $819.63 \pm 129.59$ | $2.35 \pm 0.10$ | $3,759.92 \pm 10.52$ | $1,139.23 \pm 44.35$ |
| Asia | 773 | 3,911 | 2,409,160 | 23,406 | $10.12 \pm 0.64$ | $3,116.64 \pm 281.14$ | $2.99 \pm 0.07$ | $4,089.57 \pm 4.16$ | $1,338.13 \pm 19.76$ |
| Australia | 288 | 567 | 178,447 | 2,499 | $3.94 \pm 0.39$ | $619.61 \pm 132.87$ | $2.20 \pm 0.11$ | $3,450.12 \pm 10.14$ | $810.18 \pm 39.03$ |
| Europe | 602 | 5,188 | 1,907,980 | 25,587 | $17.24 \pm 1.07$ | $3,169.40 \pm 391.04$ | $2.47 \pm 0.07$ | $2,410.24 \pm 4.05$ | $1,250.84 \pm 12.11$ |
| North America | 1,006 | 4,289 | 2,344,470 | 20,593 | $8.53 \pm 0.62$ | $2,330.48 \pm 245.49$ | $2.40 \pm 0.08$ | $3,386.44 \pm 2.66$ | $1,150.44 \pm 15.67$ |
| South America | 299 | 762 | 380,348 | 3,984 | $5.10 \pm 0.44$ | $1,272.07 \pm 154.33$ | $2.61 \pm 0.10$ | $2,719.93 \pm 6.84$ | $793.98 \pm 27.81$ |
| World | 3,237 | 18,125 | 7,440,880 | 94,644 | $11.20 \pm 0.42$ | $2,298.70 \pm 127.80$ | $2.61 \pm 0.04$ | $8,678.58 \pm 1.92$ | $1,734.64 \pm 14.58$ |

**FIG S1 - BEHAVIOR OF THE CRITICAL COOPERATION RANGE IN RANDOM NETWORKS**

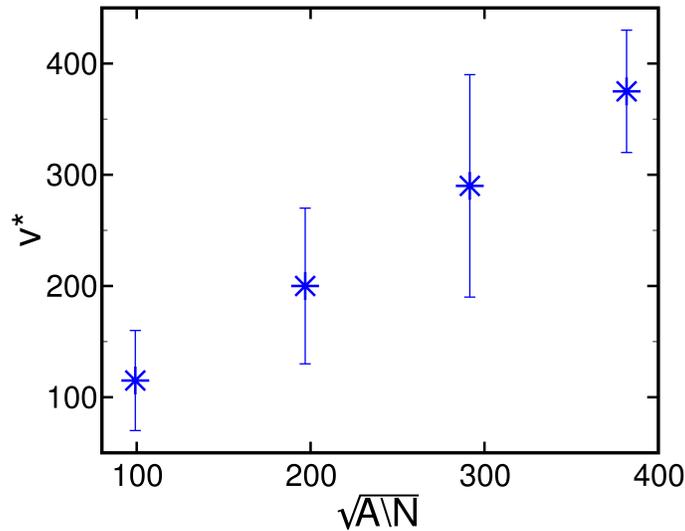

FIG. S1. The critical cooperation $v^*$ correlated with $\sqrt{A/N}$, where $A$ is the total area in which $N$ nodes are embedded, in artificial networks. Data is based on artificially generated random networks, similar to the ones used for the distance-decay model in Fig. 5 of the main text, but with links randomly assigned without any bias. Each point represent the average over 100 randomly generated networks of 500 nodes. A total of $10^4$ tentative geo swaps are executed for several cooperation range values. The value of $v^*$ is selected as the point with highest variance, with error bars representing the values where the variance is equal to $0.75\sigma^2(v^*)$.

**FIG S2 - DISTANCE RATIO IN THE MODEL**

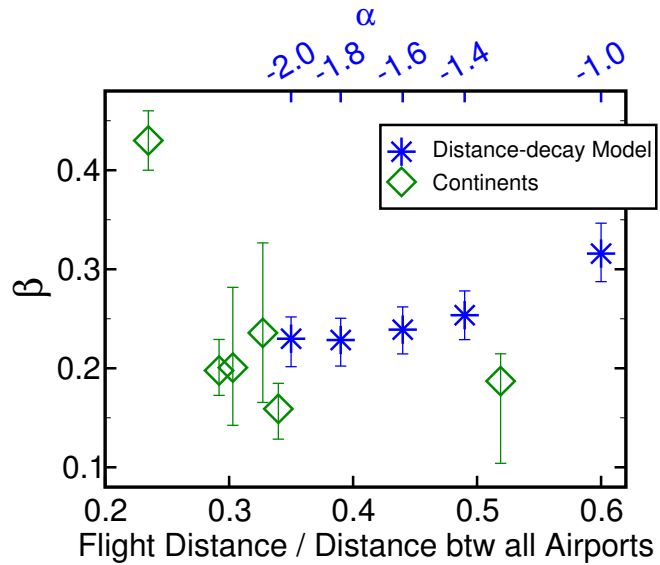

FIG. S2. Distance-decay model reproduces the same $\beta$ for different ratios between the average distance traveled by the flights (link length) and the average geographical distance between two airports of the WAN. Plot shows the impact on $\beta$ for the distance-decay model with different values of $\alpha$ (blue) in comparison with data for the continents (green). The data used is the same from Fig. 4 (continents) and Fig. 5b (distance-decay model) in the main text.

**FIG S3 - FINITE SIZE SCALING OF RANDOM NETWORKS**

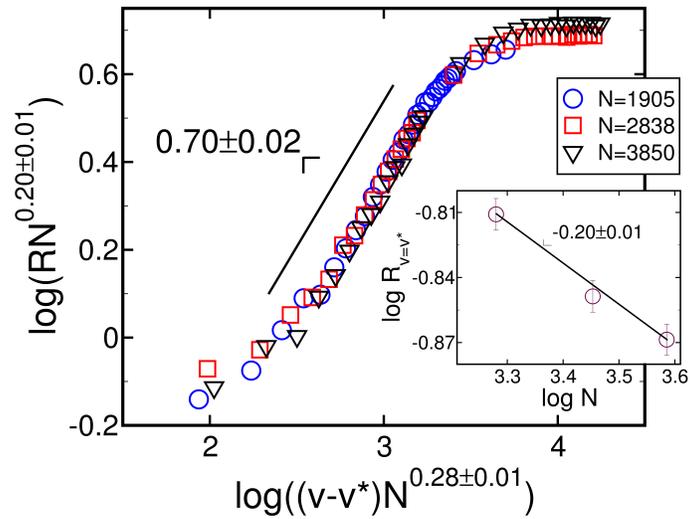

FIG. S3. Data collapse of the robustness evolution for $v - v^* > 0$ after successive applications of the geo swap in random networks. Curves in the main panel represent each continent scaled with $1/\nu = 0.28$ and $\beta/\nu = 0.20$. The inset shows the size dependence of $R$ at $v = v^*$, scaling as $R \sim N^{-\frac{\beta}{\nu}}$, with $\beta/\nu = -0.20 \pm 0.01$, where $N$ is the total number of nodes. Data is based on artificially generated random networks, similar to the ones used for the distance-decay model in Fig. 5 of the main text, but with links randomly assigned without any bias. A total of $10^4$ tentative geo swaps are executed for several cooperation range values. Each point represents the average over 200 samples, with symbols being larger than standard deviation in the main panel. The critical cooperation range is defined as $v^* = 240 \pm 10$.

FIG S4 - CHARACTERISTICS OF THE OPTIMIZED AIRPORT NETWORKS

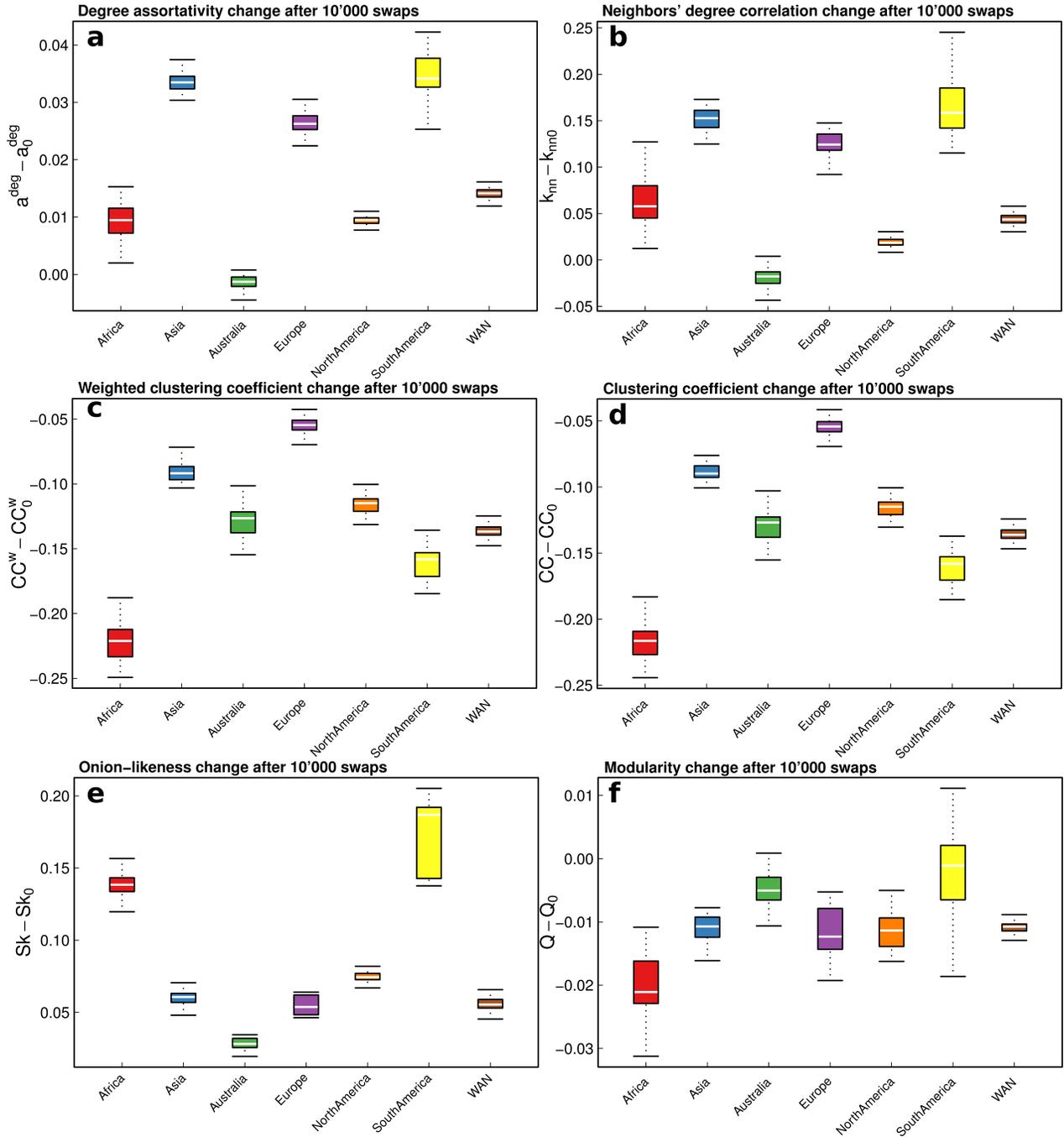

FIG. S4. Changes on networks characteristics after successive geo swaps with $v = v^*$. The strategy makes the networks more assortative, more onion-like, but also more random, as clustering coefficient and modularity decrease. Plots show the change of several features of the airport networks after $10^4$ tentative geo swaps. **a**, Degree assortativity ($a^{deg}$) [2]. **b**, Neighbors' degree correlation ($k_{nn}$) [3]. **c**, Weighted clustering coefficient ($CC^w$) [4], weighted by the number of passengers per airport. **d**, Clustering coefficient ($CC$). **e**, Onion-likeness ($Sk$) [5]. **f**, Modularity ($Q$) [6]. The subscript 0 represents the value of the value of the feature without any optimization. Box plots are used to represent the quantities computed for 100 networks, according to: lower whisker (horizontal trace below and on top of the box) for the lowest observation still within 1.5 IQR of the lower quartile (25% percentile of the distribution), bottom of the box for the lower quartile, white trace for the median, top of the box for the upper quartile, and upper whisker for the highest value still within 1.5 IQR of the upper quartile.

**FIG S5 - FLIGHT DISTANCE DISTRIBUTION**

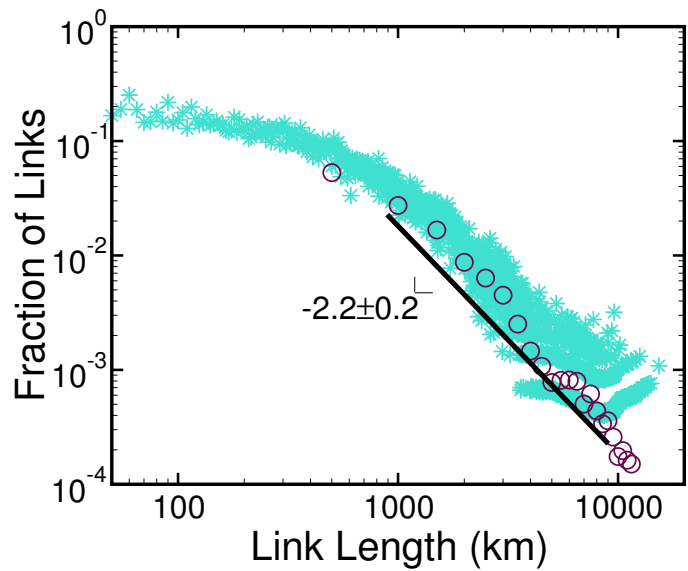

FIG. S5. Flight distance distribution (link length) of the WAN. The empirical distribution of flight length shows a power law decay with exponent $-2.2 \pm 0.2$. Both raw (turquoise) and binned (purple) data are shown in the plot. For simplicity, an exponential cutoff is not considered.